\documentclass[aps,prl,twocolumn,floatfix,superscriptaddress]{revtex4-1}

\usepackage{amsmath}
\usepackage{amssymb}
\usepackage{graphicx}
\usepackage{physics}
\usepackage{float}
\usepackage{dsfont}
\usepackage{braket}
\usepackage[colorlinks, linkcolor=blue, citecolor=blue, urlcolor=blue, breaklinks=red]{hyperref}

\newcommand{\Ccal}{\mathcal{C}}

\newcommand {\sectiontitle}[1]{\textit{#1.---}}

\begin{document}

\title{Generating long-lived entangled states with free-space collective spontaneous emission}

\author{Alan C. Santos}
\email{ac\_santos@df.ufscar.br}
\affiliation{Departamento de F\'isica, Universidade Federal de S\~{a}o Carlos, Rod.~Washington Lu\'is, km 235 - SP-310, 13565-905 S\~{a}o Carlos, SP, Brazil}

\author{Andr\'e Cidrim}
\email{andrecidrim@gmail.com}
\affiliation{Departamento de F\'isica, Universidade Federal de S\~{a}o Carlos, Rod.~Washington Lu\'is, km 235 - SP-310, 13565-905 S\~{a}o Carlos, SP, Brazil}

\author{Celso Jorge Villas-Boas}
\email{celsovb@df.ufscar.br}
\affiliation{Departamento de F\'isica, Universidade Federal de S\~{a}o Carlos, Rod.~Washington Lu\'is, km 235 - SP-310, 13565-905 S\~{a}o Carlos, SP, Brazil}

\author{Robin Kaiser}
\email{robin.kaiser@inphyni.cnrs.fr}
\affiliation{Universit\'e C\^ote d'Azur, CNRS, Institut de Physique de Nice, 06560 Valbonne, France}

\author{Romain Bachelard}
\email{bachelard.romain@gmail.com}
\affiliation{Departamento de F\'isica, Universidade Federal de S\~{a}o Carlos, Rod.~Washington Lu\'is, km 235 - SP-310, 13565-905 S\~{a}o Carlos, SP, Brazil}

\begin{abstract}
Considering the paradigmatic case of a cloud of two-level atoms interacting through common vacuum modes, we show how cooperative spontaneous emission, which is at the origin of superradiance, leads the system to long-lived entangled states at late times. These subradiant modes are characterized by an entanglement between all particles, independently of their geometrical configuration. While there is no threshold on the interaction strength necessary to entangle all particles, stronger interactions lead to longer-lived entanglement. 
\end{abstract}

\date{\today}

\maketitle

\sectiontitle{Introduction}Entangled states, apart from their interest for fundamental physics~\cite{Zeilinger1999}, are now turning into crucial resources for quantum information science, in particular for secure quantum communication~\cite{Gisin2002} and quantum metrology~\cite{Polino2020}. Once created, these states suffer from decoherence, due to their inevitable coupling to the environment. This can be circumvented by different strategies, such as resorting to decoherence-free subspaces~\cite{Lidar1998,Kwiat2000}, the quantum Zeno effect~\cite{Facchi2004,Maniscalco2008}, or weak measurements~\cite{Kim2011}. Alternatively, it has been proposed to engineer locally the decoherence to reach a target entangled state~\cite{Diehl2008,Kraus2008,Verstraete2009}. Yet, the required control of each decay channel makes it an unpractical and non-scalable solution.

Differently, we here show that collective spontaneous emission, which arises from dipole-dipole interactions, naturally leads to the formation of long-lived entangled states. These interactions present the peculiarity of being composed of both a coherent and a dissipative part~\cite{Lehmberg1970}, often referred to virtual and real photons, respectively. The hallmark of collective spontaneous emission is superradiance~\cite{Dicke1954}, which was initially studied in the context of a cascade, as the system decays from a fully excited toward its ground state, passing through a series of symmetric entangled states~\cite{Gross1982}. Note that the entangled nature of the visited states was later questioned, since semi-classical approaches exhibit very similar features~\cite{Arrechi1972,MacGillivray1976}.

\begin{figure}
\includegraphics[width=.485\textwidth]{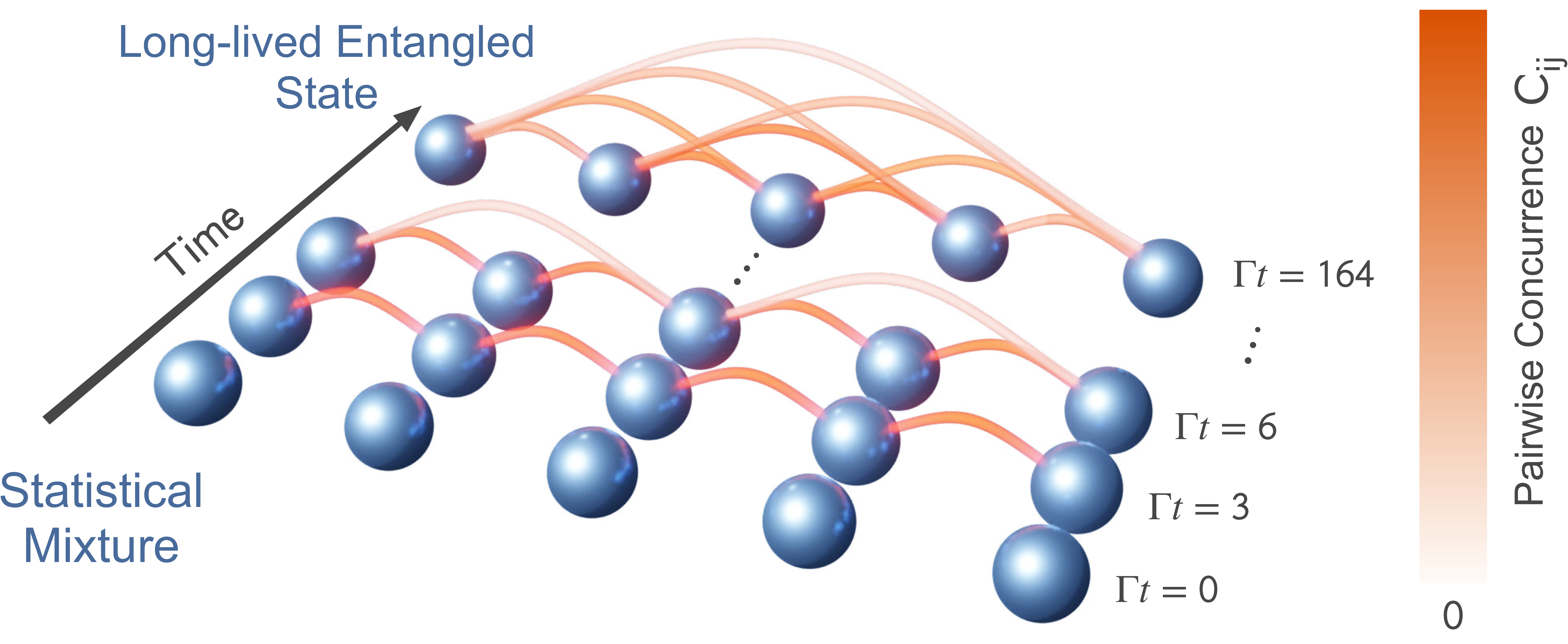}
\caption{Propagation of the entanglement in a chain of $N=5$ atoms, starting from a statistical mixture [see Eq.~\eqref{eq:mix}], and evolving to an all-to-all entangled state. Simulations realized for a lattice step $kd=\pi/2$, with polarization of the atoms perpendicular to the atomic chain. The colorbar is rescaled at each time [with $\max(C_{ij})=0, \ 0.03, 0.08$, and $0.007$ from earlier to later times].}
\label{Fig:scheme}
\end{figure}

Dipole-dipole interactions have attracted a renewed attention over the past years, with a large focus on ``single-excitation'' collective processes. In this weak-driving regime,  super- and subradiance were re-examined~\cite{Guerin2016,Roof2016,Araujo2016,Das2020}, and the superflash~\cite{Kwong2014} and collective steady-state shifts were also observed~\cite{Ido2005,Scully2009,Keaveney2012,Javanainen2014,Meir2014,Roof2016,Jennewein2016,Jennewein2016b,Jenkins2016,Bromley2016,Peyrot2018}. These studies were restricted to the single-excitation states, which allows to explore a reduced portion of the Hilbert space (of size $N+1$ instead of $2^N$, for $N$ two-level emitters). However, only the use of a single-photon source (and thus quantum light) can guarantee an at-most-single-excitation state~\cite{deOliveira2014}, or the presence of specific mechanisms such as excitation blockade~\cite{Cidrim2020,Williamson2020}. Indeed, collective effects were shown to challenge the notion of ``weak drive'', since the narrow-linewidth collective modes (that is, the subradiant modes) present a nonlinear reaction to the drive even at very low saturation parameter~\cite{Williamson2020a,Cipris2021}.

In this work, we show how at-most-single-excitation states can be created out of a statistical mixture by collective spontaneous emission, without resorting to strong energy shifts to address specific modes with an appropriate drive frequency (such as for blockade-like mechanisms) or to single-photon pulses with specific phase patterns. The longest-lived modes are single-excitation ones, which make them hold most of the excited population; indeed highly excited modes decay faster than the single-excited states and this process, interestingly, includes decay channels into long-lived single-excitation subradiant states.

These long-lived modes are characterized by a finite concurrence, which can survive over timescales hundreds of times larger than the single atom excited state lifetime, see Fig.~\ref{Fig:scheme}. We first investigate linear regular chains, before showing that three-dimensional disordered clouds present the same features. Remarkably, all atoms always get entangled altogether, even for vanishing interactions, although stronger interactions guarantees that the associated concurrence survives longer. These entangled states leave a direct signature in the radiated light, whose equal-time second-order optical coherence vanishes at longer times. Our work paves the way toward preparation of dissipatively-induced globally-entangled states, even in large atomic systems.

\sectiontitle{Collective spontaneous emission}Let us consider an ensemble of $N$ two-level atoms, at positions $\mathbf{r}_j$, with a transition between their ground and excited states $\ket{g}$ and $\ket{e}$ characterized by its transition frequency $\omega_a=kc=2\pi c/\lambda$, linewidth $\Gamma$, and rising (lowering) operators $\hat\sigma_j^{+}$ ($\hat\sigma_j^{-}$). The dipole dynamics is obtained from a master equation $\dot{\hat{\rho}} = -i [\hat H, \hat\rho] + \mathcal{L}(\hat \rho)$, with a coherent (Hamiltonian) contribution~\cite{Stephen1964, Lehmberg1970, Friedberg1973}:
\begin{align}
\hat H &= -\Delta \sum_j  \hat \sigma_j^+  \hat \sigma_j^- +  \frac{1}{2} \sum_j \left( \Omega e^{i\mathbf{k}\cdot \mathbf{r}_j} \hat \sigma_j^+  + h.c. \right)  \nonumber \\
&+\sum_{j,m\neq j} \Delta^{jm} \hat \sigma_j^+ \hat \sigma_m^-, \label{eq:H_dd}
\end{align} 
written in the rotating frame of the pump driving frequency, and a dissipative (Lindbladian) component
\begin{align}
\mathcal{L}(\hat \rho) &= \frac{1}{2}\sum_{j,m} \Gamma^{jm} \left(2\hat \sigma_j^- \hat \rho \hat \sigma_m^+ - \{ \hat \sigma_m^+ \hat \sigma_j^-,\hat \rho \}\right) \label{eq:diss}. 
\end{align}

\noindent We assume point-like dipoles, whose associated Green's tensor is given by $\mathbf{G}_{jm}\equiv\mathbf{G}(\mathbf{r}_{jm})=\frac{3\Gamma}{4}\frac{e^{ikr_{jm}}}{(kr_{jm})^3}\Bigl[\Large(k^2r_{jm}^2+ikr_{jm}-1\Large)\mathds{1}_3-\Large(k^2r_{jm}^2+i3kr_{jm}-3\Large)\frac{\mathbf{r}_{jm}\mathbf{r}_{jm}^{T}}{r_{jm}^2}\bigr]$ for $j\neq m$, where $\mathbf{r}_{jm}\equiv \mathbf{r}_j - \mathbf{r}_m$, and $\mathbf{G}_{jj}=i\frac{\Gamma}{2}\mathds{1}_3$ for the single-atom term. Thus the incoherent coupling term is $\Gamma^{jm}\equiv \hat{\epsilon}_j^{*}\cdot 2\mathrm{Im}\left\{\mathbf{G}_{jm}\right\}\cdot\hat{\epsilon}_m$, while the excitation-exchange term is $\Delta^{jm}\equiv-\hat{\epsilon}_j^{*}\cdot \mathrm{Re}\left\{\mathbf{G}_{jm}\right\}\cdot \hat{\epsilon}_m$, with $\hat{\epsilon}_j$ the polarization of the $j$-th dipole, here chosen as $\hat\epsilon_j=\hat z$. These coupling terms stem from the interaction of the dipoles through common radiation modes, and they are at the origin of the superradiant/subradiant decay and collective energy shifts. The purpose of the classically-treated monochromatic plane wave in the Eq. \eqref{eq:H_dd}, with Rabi frequency $\Omega e^{i\mathbf{k}\cdot \mathbf{r}}$ and detuned from the atomic transition by $\Delta$, is only to prepare the system, and it is turned off at $t=0$ to study the decay dynamics.

\begin{figure}
\includegraphics[width=.45\textwidth]{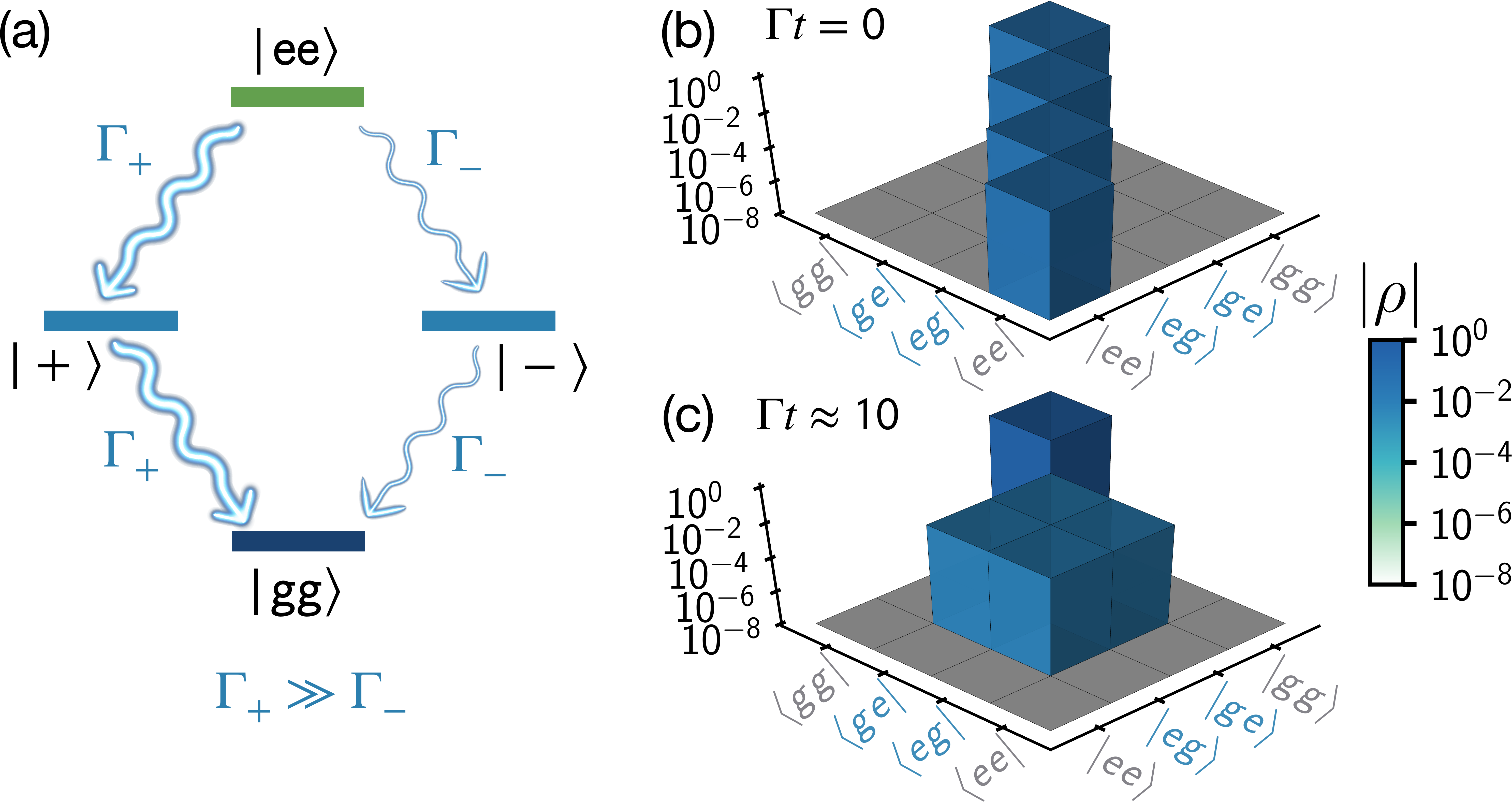}
\caption{(a) Energy levels for a pair of atoms, with the superradiant (subradiant) decay denoted by a thick (thin) arrow. (b-c) Density matrix of the system at the initial time [(b) $\Gamma t=0$], when the system is in the statistical mixture $\hat{\rho}_\textrm{mix}$, and at a late time [(c) $\Gamma t=10$], when the system decayed toward an entangled state of the form $\hat\rho\approx(1-\epsilon)\ket{gg}\bra{gg}+\epsilon\ket{-}\bra{-}$. Atoms distant of $d=0.1/k$; energy shifts are not represented, since they do not play a role in the decay process.}
\label{fig:N2}
\end{figure}

\sectiontitle{Two-atom case}Let us first discuss the case of a pair of close atoms ($N=2$), since it offers an intuitive picture of how a long-lived entangled state can be generated by collective spontaneous decay~\cite{You21}, even when starting from a statistical mixture. In Fig.~\ref{fig:N2}(a) we illustrate the different states composing the Hilbert space, with their decay rate.  The excited states are the fully-excited one $\ket{ee}$, which decays at rate $2\Gamma$, and the symmetric and antisymmetric single-excitation states, $\ket{\pm}=(\ket{eg}\pm\ket{ge})/\sqrt{2}$, which decay at rate $\Gamma_\pm$ toward the ground state $\ket{gg}$. For two strongly interacting atoms ($r_{12}\ll\lambda$), $\Gamma_+\approx2\Gamma$ and $\Gamma_-\ll\Gamma$. Consequently, when the drive is switched off, the population of states $\ket{ee}$ and $\ket{+}$ decay to zero on a timescale $1/\Gamma$, whereas the (single-excitation) antisymmetric state holds its population over times $1/\Gamma_-\gg 1/\Gamma$. The latter time can be arbitrarily large, as the two atoms are approximated and assuming there is no additional source of decoherence. 

Let us first consider a statistical mixture, such as created by a strong drive, as initial state, since it has been shown to be an efficient scheme to populate efficiently long-lived states~\cite{Cipris2021}: 
\begin{equation}
\hat\rho_\textrm{mix}=\bigotimes_{j=1}^{N}\left( \frac{\ket{g_j}\bra{g_j}+\ket{e_j}\bra{e_j}}{2}\right).\label{eq:mix}
\end{equation}
For $N=2$ atoms, the antisymmetric (subradiant) mode $\ket{-}$ then holds one fourth of the population, just like the other states. According to the previous reasoning, a few units of $1/\Gamma$ after the pump is switched off, the system is found in an entangled state:  $\hat\rho_a\approx \frac{3}{4}\ket{gg}\bra{gg}+\frac{1}{4}\ket{\psi_-}\bra{\psi_-}$. This is illustrated in Fig.~\ref{fig:N2}(c), where the density matrix of the system is presented at $\Gamma t=10$ after switch-off. We hereafter quantify the entanglement using the concurrence as proposed by Hill and Wootters~\cite{Hill:97,Wootters:98}. It is defined as $\Ccal(\hat\rho) \equiv \max \{ 0, \lambda_{1} - \lambda_{2} - \lambda_{3} - \lambda_{4}\}$, with $\lambda_{n}$ the eigenvalues of the matrix $\hat R\!=\! ( \hat\rho^{1/2}\hat{\tilde{\rho}}\hat\rho^{1/2} )^{1/2}$ in decreasing order ($\lambda_{n}\geq\lambda_{n+1}$); we have here introduced $\hat{\tilde{\rho}}\equiv(\hat\sigma_{y}\otimes \hat\sigma_{y}) \hat\rho^{*} (\hat\sigma_{y}\otimes \hat\sigma_{y})$, with $\hat\rho^{*}$ the complex conjugate of $\hat\rho$ taken in the atomic basis $\{\ket{g},\ket{e}\}$. The concurrence for the state $\hat\rho_a$ is $\mathcal{C}(\hat\rho_a)\approx 0.25$ at a time $\Gamma t=10$.

\sectiontitle{Many-atom case}Moving to the many-atom case, one faces the usual challenge of the exponential growth of the Hilbert space with $N$, which is aggravated by the fact that the collective spontaneous emission is described by a density matrix and its associated Lindblad superoperator $\mathcal{L}$~\cite{Breuer2007}. The presence of multiply-excited subradiant states may actually prevent the system to reach an entangled state, as the superposition of Dicke (entangled) states is not necessarily entangled. Nevertheless, the longest-lived modes are actually encountered in the single-excitation manifold~\cite{Asenjo2017,Cipris2021}, with the lifetime of the {\it longest-lived} $n$-excitation state scaling as $(n/N)^{-3}$. Interestingly, this scaling actually guarantees that the population from doubly- and higher-excited states decays much faster than that of the longest-lived single excitation one, so the system is bound to decay toward a mixture between ground and single-excitation states.

Beyond this qualitative argument, the variety of states and decay channels between them calls for a numerical approach, which we conduct using exact simulations~\cite{Johansson2012,Johansson2013}, at first for a linear regular chain of $N$ emitters with step $d$. The system is prepared in the statistical mixture $\hat\rho_\textrm{mix}$ introduced in Eq.~\eqref{eq:mix}, and the strong pump is again switched off at time $t=0$. We monitor the generation of entanglement by computing the average pair entanglement $\mathcal{C}_\textrm{avg}=[\sum_{j,m\neq j}\mathcal{C}_{jm}]/N(N-1)$, based on the  concurrence for pairs of atoms $\mathcal{C}_{jm}=\Ccal(\hat\rho_{jm})$, where $\hat\rho_{jm}$ is the reduced density matrix for the atoms $j$ and $m$.

\begin{figure}
\includegraphics[width=.49\textwidth]{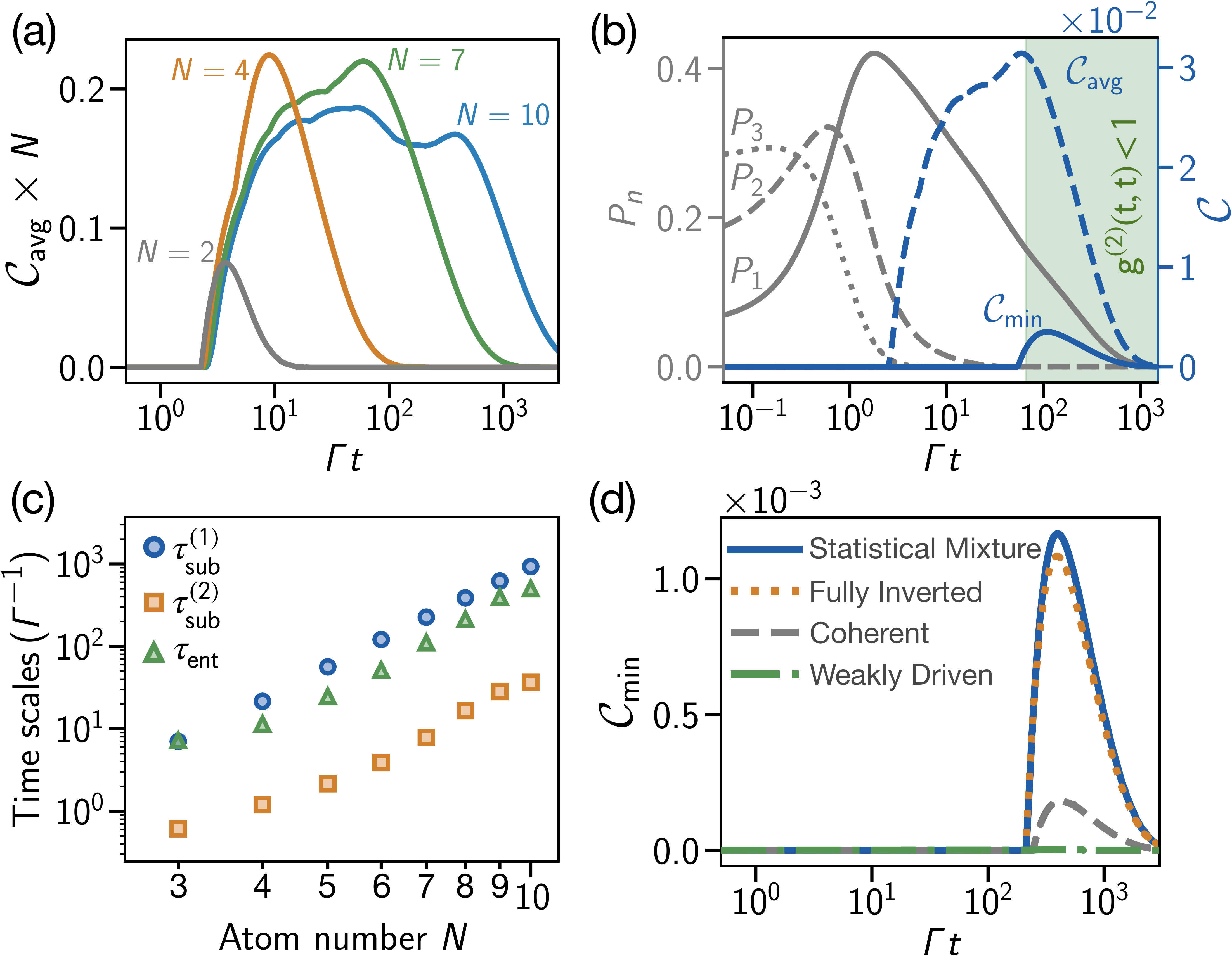}
\caption{(a) Average concurrence multiplied by $N$, as a function of the time and for chains of different atom numbers ($N=2,\ 4,\ 7,\ 10$) and with spacing $kd=\pi/2$. (b) Dynamics of the $n$-excitation state populations $P_n$ (for $n=1,\ 2,\ 3$ and $N=7$ atoms), along with average $\mathcal{C}_\mathrm{avg}$ and minimum concurrence $\mathcal{C}_\mathrm{min}$, as a function of time. The green shaded area indicates the time interval where $g^{(2)}(t,t)<1$. (c) Lifetime $\tau_\mathrm{sub}^{(n)}=1/\Gamma_\mathrm{sub}^{(n)}$ of the longest-lived state with $n$ excitations, for $n=1,\ 2$, and time $\tau_\textrm{ent}$ at which the peak of $\mathcal{C}_\textrm{min}$ is reached, as a function of the atom number $N$. (d) Minimum concurrence $\mathcal{C}_\mathrm{min}$ for different initial states: statistical mixture \eqref{eq:mix} (solid blue curve), fully inverted (dotted orange curve), coherent $\psi_\mathrm{coh}=\bigotimes_j( \ket{g_j}+\ket{e_j})/\sqrt{2}$ (dashed gray curve) and weak-drive steady-state for $\Omega=0.1\Gamma$ (dash-dotted green curve).}
\label{fig:C}
\end{figure}

The dynamics of the average concurrence is presented in Fig.~\ref{fig:C}(a), where the peak value of the concurrence $\mathcal{C}_\textrm{avg}$ is observed to scale roughly as $1/N$. This scaling stems from the concurrence of the Dicke states themselves. Indeed, calling $\ket{\psi_{1e}}=(1/\sqrt{N})\sum_j e^{i\phi_j}\ket{g...e_j...g}$ the single-excitation Dicke state, one can show that a state of the form $\hat{\rho}_\epsilon=(1-\epsilon)\ket{g...g}\bra{g...g}+\epsilon\ket{\psi_{1e}}\bra{\psi_{1e}}$ has an average concurrence $2\epsilon/N$, for any set of phases $\{\phi_j\}$~\cite{SupMat}. Hence, the observed value $\mathcal{C}_\textrm{avg}\approx0.2/N$ suggests that a fraction of $\sim 10\%$ of the system state decays toward the single-excitation most-subradiant state, at a time when higher-excited states have a negligible population. 

This is confirmed by a closer analysis of the evolution of the populations $P_n$ of the $n$-excitation states:
\begin{equation}
P_{n}=\mathrm{Tr}\left[\left(\sum_{\mathcal{P}}\otimes_{j=1}^n \dyad{e_j}{e_j}\otimes_{m=n+1}^N\dyad{g_m}{g_m}\right)\hat\rho\right].
\end{equation}
In Fig.~\ref{fig:C}(b), the concurrence $\mathcal{C}_\textrm{avg}$ reaches a late peak when states with $n\geq2$ excitations have depleted, and only the single-excited and ground states remain populated. This peak is associated with a value $P_1\approx 0.15$, consistent with the above hypothesis of the system being in a mixture between ground and single-excitation Dicke state. We note that one may also seek signatures of the long-lived states population in the transmission of a weak pulse through the atomic sample, as it was done for few-emitter systems~\cite{Slodicka2010,Rossatto2013}.

Nevertheless, the average concurrence $\mathcal{C}_\textrm{avg}$ provides limited information, in the sense that it is non-zero already when only a pair of atoms is entangled. A stronger measure is the minimum concurrence $\mathcal{C}_\textrm{min}=\textrm{min}_{(j,m)}\mathcal{C}_{jm}$, taken over all pairs of atoms $j$ and $m$, which is nonzero only if {\it all} pairs of atoms are entangled with each other. This minimum concurrence departs from zero only at late times, when all multiply-excited states have negligible populations, see Fig.~\ref{fig:C}(b). Hence, a global entanglement between all  atoms is reached only at the latest time, after collective spontaneous emission has driven the system toward a mixture between a collective (Dicke) single-excitation state and the ground state.

The picture of the maximum of global entanglement being reached after multiply-excited states have died out is confirmed by comparing, for different atom numbers, the lifetime of the longest-lived single- and double-excited states ($\tau_\textrm{sub}^{(n)}$ with $n=1,\ 2$) with the time $\tau_\textrm{ent}$ at which the peak of $\mathcal{C}_\textrm{min}$ is reached. As presented in Fig.~\ref{fig:C}(c), this global entanglement is maximum at times when double-excited states have vanished, yet single-excited states have not, that is: $\tau_\textrm{sub}^{(2)}<\tau_\textrm{ent}<\tau_\textrm{sub}^{(1)}$. A remarkable point is that this global entanglement actually never dies, provided there is no additional source of decoherence. Indeed, at late times the system is essentially in the at-most-single-excitation state $\hat{\rho}_\epsilon$ introduced before, with $\epsilon\propto \exp(-t/\tau^{(1)}_\textrm{sub})$, so the minimum concurrence decays exponentially at the slow rate $1/\tau^{(1)}_\textrm{sub}$.

Entanglement is thus generated by cooperative decoherence, starting from a statistical mixture where no correlations between the particles are present. This rises the question of the optimal state to start the decay process from, in order to achieve larger entanglement values. The case of an initially fully-inverted system, which was the configuration initially proposed by Dicke to study the superradiant cascade~\cite{Dicke1954}, is actually slightly less efficient at generating an all-entangled state, as shown in Fig.~\ref{fig:C}(d). The product state $\psi_\mathrm{coh}=\bigotimes_j( \ket{g_j}+\ket{e_j})/\sqrt{2}$, where each atom is in a coherent state, leads to a much weaker entanglement, due to the fact that, differently from the statistical mixture, it has a very reduced projection on subradiant states (for $N=2$ atoms, this projection is initially zero). Finally, a weak drive (a resonant plane-wave with Rabi frequency $\Omega=0.1\Gamma$) is even less efficient at creating entanglement, due to the poor coupling of subradiant states to plane-waves in the weak-drive regime. Hence, because the statistical mixture \eqref{eq:mix} is the most efficient at populating long-lived state by decoherence~\cite{Cipris2021b}, it also generates most efficiently a long-lived globally-entangled atomic state.

\begin{figure}
\includegraphics[width=.48\textwidth]{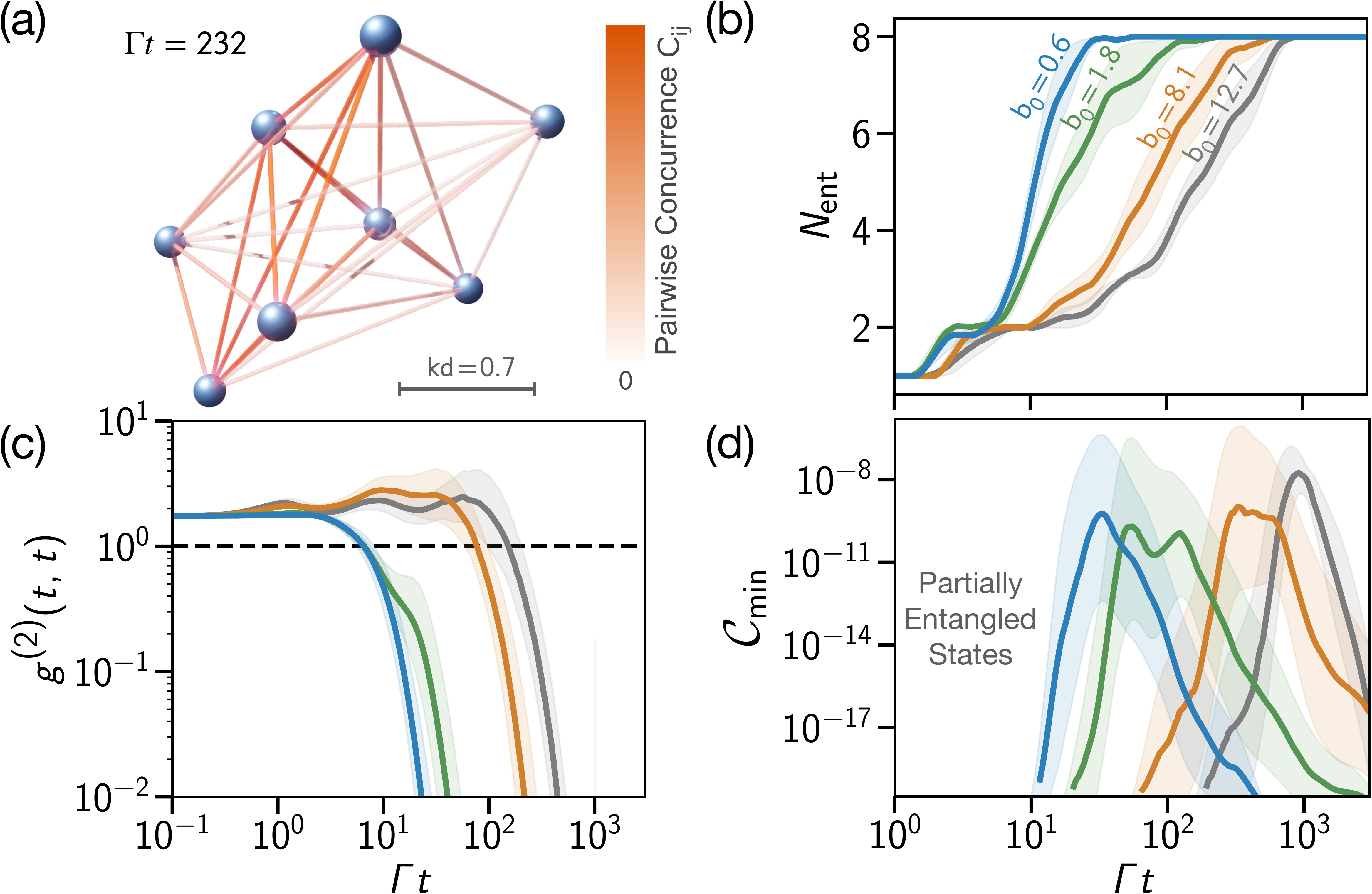}
\caption{(a) Illustration of the entanglement generated by collective spontaneous emission in a disordered cloud of $N=8$ atoms, at a time $t=232\,\Gamma^{-1}$ after starting the decay process from state \eqref{eq:mix} [with $\max(C_{ij})\approx8\times 10^{-8}$]; (b) Number of entangled atoms (see text) as a function of time, (c) Equal-time second-order coherence of the radiated light, and (d) Minimum concurrence $\mathcal{C}_\textrm{min}$, as a function of time, and for clouds of $N=8$ atoms and of different optical depths. The shaded areas represent the fluctuations encountered over the 10 realizations, for a homogeneous-density cloud with radius such that $b_0=2N/(kR)^2$ [$kR\approx 5.2,\ 3,\ 1.4$ and $1.1$ for the increasing values of $b_0$ presented].}
\label{fig:disordered}
\end{figure}

\sectiontitle{Disordered clouds}Let us now discuss the case of disordered clouds which, despite the absence of specific interference patterns, are known to hold superradiance and subradiance, both arising from cooperative spontaneous emission. As mentioned previously, the large difference in lifetimes between long-lived single- and double-excited states, illustrated in Fig.~\ref{fig:C}(c) for linear chains~\cite{Asenjo2017}, has also been reported for disordered samples~\cite{Cipris2021}. This leads to the generation of an all-entangled state at late times, as before, although the distribution of pair concurrence no longer exhibits a simple pattern, see Fig.~\ref{fig:disordered}(a).

As for the linear chains (see Fig.~\ref{Fig:scheme}), the entanglement by pairs is created progressively, until it spreads over all the system. In Fig.~\ref{fig:disordered}(b), we present the number of entangled atoms $N_\textrm{ent}$, or size of the entangled cluster, defined as the size of the largest subset (or cluster) of particles in which all pairs are entangled. One observes that independently of the resonant optical thickness $b_0=2N/(kR)^2$ ($R$ the cloud radius), which is a measure of the cooperativity in disordered clouds~\cite{Guerin2017b}, the system decays to a collective state where all atoms are entangled, that is, $N_\textrm{ent}=N$ at late times. 

It may seem surprising that even for weak interactions (small values of $b_0$), all pairs of atoms develop entanglement. The explanation can be found in the values of the minimum concurrence associated with these late-time states: weaker interactions come with shorter timescales for the concurrence. Indeed, since the creation of entanglement relies on the difference between the lifetimes of single- and double-excited long-lived states, the fact that both tend to $1/\Gamma$ for vanishing interactions implies that the time over which the entanglement is substantial vanishes as well. This point is illustrated in Fig.~\ref{fig:disordered}(d), where the increasing values of $b_0$ are characterized by larger timescales for observing a finite minimum concurrence. Note that we have used, for the disordered case, the scalar model for dipole-dipole interactions. Indeed, the near-field terms strongly affect the subradiant decay, a phenomenon which can interpreted as van der Waals dephasing~\cite{Cipris2021b}, and it here leads to a reduction of the entanglement at higher densities~\cite{SupMat}.

\sectiontitle{Signature in the scattered light}Interestingly, this creation of collective single-excitation states leaves a direct signature in the light radiated by the cloud. Let us introduce the equal-time second-order optical coherence
\begin{equation}\label{eq:g2}
g^{(2)}(t,t)\equiv \frac{\langle \hat{E}^{-}(t)\hat{E}^{-}(t)\hat{E}^{+}(t)\hat{E}^{+}(t)\rangle}{\langle \hat{E}^{-}(t)\hat{E}^{+}(t) \rangle^2},
\end{equation}
which reflects the capacity of the system to emit two photons at a time $t$. Here, $\hat{E}^\dagger\propto\sum_j e^{-ik\hat{n}.\mathbf{r}_j}\sigma_j^-$ refers to the radiated field in a direction $\hat{n}$, in the far-field limit. For a state with at most one excitation (that is, without contribution from multiply-excited states), one has $g^{(2)}=0$. In Fig.~\ref{fig:disordered}(d), one observes that the $g^{(2)}(t,t)$ starts for $t<1/\Gamma$ slightly below $2$, as expected from a system composed of several independent emitters [the system is initially in the statistical mixture given in Eq.~\eqref{eq:mix}]. Then it undergoes a burst (more visible for larger values of $b_0$), as the system decays superradiantly toward lowly-excited states. Finally, the $g^{(2)}(t,t)$ goes below unity at the same time as the minimal concurrence rises above zero, that is, when the system can be considered in a superposition between ground and single-excited collective state. Thus, the below-unity $g^{(2)}(t,t)$ is a signature of the at-most-single-excitation nature of the state, whereas its long lifetime reflects its collective (subradiant) nature. This is analogous to the blockade mechanism, where the population-population correlations witness the single-excitation nature of the state, whereas the accelerated Rabi oscillations reveal its collective nature~\cite{Gaetan2009}. Note that we have checked that the detection of at most on excitation in the state by the intensity-intensity correlation function is, at late times, unaffected by the finite temporal resolution of the photodetector~\cite{SM}. This stems from the fact that the at-most-single-excitation nature of the state remains true at any later time.

\sectiontitle{Conclusion}We have shown how cooperative spontaneous emission drives the system toward a state where all atoms are pair-entangled, at late times when it is in a superposition between ground and single-excitation states. While this mechanism is valid for arbitrary interaction strength (here determined by the distance between the atoms), a larger cooperativity promotes longer-lived entanglement. Furthermore, because it does not rely on specific interference patterns between the atoms, the phenomenon is equally present in disordered systems. 

The creation and propagation of entanglement by decoherence in this long-range interacting system leads to several intriguing questions, such as whether the excitation-exchange (Hamiltonian) or the spontaneous-emission (Lindbladian) dynamics is more efficient to propagate quantum correlations~\cite{Poulin2010,Sweke2019,Tran2021}, or if spin squeezing may also be produced using decoherence~\cite{Qu2019}.

\sectiontitle{Note}During the writing of this manuscript, we have become aware of a related work on the generation of entanglement by decoherence in 1D systems~\cite{You21}.

\sectiontitle{Acknowledgments} A.C.S., A.C., C.J.V.-B. and R.B. are supported by the S\~ao Paulo Research Foundation (FAPESP) through Grants No. 2019/22685-1, 2017/09390-7, 2019/13143-0, 2019/11999-5 and 2018/15554-5. R.B. received support from the National Council for Scientific and Technological Development (CNPq) Grant Nos. 313886/2020-2 and 409946/2018-4. Part of this work was performed in the framework of the European Training Network ColOpt, which is funded by the European Union (EU) Horizon 2020 program under the Marie Sklodowska-Curie action, grant agreement No. 721465. R. B. and R. K. received support from project CAPES-COFECUB (Ph879-17/CAPES 88887.130197/2017-01), and project STIC-AmSud 47221YJ/CAPES 88881.521973/2020-01. R. K. received support from the European project ANDLICA, ERC Advanced Grant Agreement No. 832219. 

\sectiontitle{Statement of equal contribution} The authors A.C.S. and A.C. contributed equally to this work.

\bibliography{BiblioCollectiveScattering.bib}

\end{document}


\title{Supplemental Material for: \\ Generating long-lived entangled states with free-space collective spontaneous emission}

\author{Alan C. Santos}
\email{ac\_santos@df.ufscar.br}
\affiliation{Departamento de F\'isica, Universidade Federal de S\~{a}o Carlos, Rod.~Washington Lu\'is, km 235 - SP-310, 13565-905 S\~{a}o Carlos, SP, Brazil}

\author{Andr\'e Cidrim}
\email{andrecidrim@gmail.com}
\affiliation{Departamento de F\'isica, Universidade Federal de S\~{a}o Carlos, Rod.~Washington Lu\'is, km 235 - SP-310, 13565-905 S\~{a}o Carlos, SP, Brazil}

\author{Celso Jorge Villas-Boas}
\email{celsovb@df.ufscar.br}
\affiliation{Departamento de F\'isica, Universidade Federal de S\~{a}o Carlos, Rod.~Washington Lu\'is, km 235 - SP-310, 13565-905 S\~{a}o Carlos, SP, Brazil}

\author{Robin Kaiser}
\email{robin.kaiser@inphyni.cnrs.fr}
\affiliation{Universit\'e C\^ote d'Azur, CNRS, Institut de Physique de Nice, 06560 Valbonne, France}

\author{Romain Bachelard}
\email{bachelard.romain@gmail.com}
\affiliation{Departamento de F\'isica, Universidade Federal de S\~{a}o Carlos, Rod.~Washington Lu\'is, km 235 - SP-310, 13565-905 S\~{a}o Carlos, SP, Brazil}

\maketitle

\appendix

\setcounter{equation}{0}
\setcounter{figure}{0}
\setcounter{table}{0}

\renewcommand{\theequation}{S\arabic{equation}}
\renewcommand{\thefigure}{S\arabic{figure}}
\renewcommand{\bibnumfmt}[1]{[S#1]}
\renewcommand{\citenumfont}[1]{S#1}

\section{Concurrence for Dicke states}

Let us call $\ket{D_{N}}$ a Dicke state from the single-excitation subspace of the Hilbert space $\Hcal$ of $N$ two-level atoms:
\begin{align}
\ket{D_{N}} = \frac{1}{\sqrt{N}} \sum_{n=1}^{N} e^{i\phi_{n}}\ket{e}_{n}\ket{g}_{m\neq n}^{\otimes N-1} ,
\end{align}
where $\phi_{n}$ are arbitrary phases. We note that since these phases can be changed by local single-atom rotations, they do not contribute to the amount of classical and quantum correlations~\cite{Nielsen:Book}. Then, for clarity, we hereafter consider $\phi_{j}=0,\ \forall j$. Let us now compute the pairwise concurrence between the $k$-th and $\ell$-th atoms, which requires the reduced density matrix for the atoms $\hat\rho_{k\ell}^{D}\!=\! \mathrm{Tr}_{\Hcal\neq\Hcal_{k\ell}}[\rho_{D}]$, where $\Hcal_{k\ell}$ the Hilbert subspace associated with the pair of atoms $(k,\ell)$. Introducing $\hat\rho_{D}=\ket{D_{N}}\bra{D_{N}}$, we write
\begin{align}
\rho_{k\ell}^{D} = \sum_{j_{r}}^{(k,\ell)} \bra{\{j_{r}\}_{r\neq(k,\ell)}} \hat\rho_{D} \ket{\{j_{r}\}_{r\neq(k,\ell)}},
\label{Ap-Eq-RhoklI}
\end{align}
with $\{\ket{\{j_{r}\}_{r\neq(k,\ell)}}\}$ the canonical basis of the Hilbert space of the remaining $N-2$ atoms, and where the symbol $\sum_{j_{r}}^{(k,\ell)}$ denotes the sum over all states $\{\ket{\{j_{r}\}_{r\neq(k,\ell)}}\}$. 
This leads to the following reduced matrix for the two atoms $k$ and $\ell$:
\begin{align}
\hat\rho_{k\ell}^{D} &= \underbrace{\frac{N-2}{N} \ket{00}\bra{00}_{k\ell} + \frac{1}{N} \left( \ket{10}\bra{10}_{k\ell} + \ket{01}\bra{01}_{k\ell} \right)}_{\text{Populations}} \nonumber \\
&+ \underbrace{\frac{1}{N} \left( \ket{10}\bra{01}_{k\ell} + \ket{01}\bra{10}_{k\ell} \right)}_{\text{Coherences}} . \label{Ap-Eq-rholk-Pure}
\end{align}
The last term in the above equation, which describes coherences, already suggests an amount of quantum correlations scaling as $1/N$, as we shall now demonstrate. Let us quantify the amount of entanglement between atoms $k$ and $\ell$ by their concurrence (see main text). To this end, we compute the eigenvalues of the matrix $\hat R=( \hat\rho^{1/2}\hat{\tilde{\rho}}\hat\rho^{1/2} )^{1/2}$ for state \eqref{Ap-Eq-rholk-Pure}, which leads to $\lambda_{1}=2/N$ and $\lambda_{n\neq1}=0$, so the concurrence can be deduced:
\begin{align}
\Ccal(\hat\rho_{k\ell}^{D}) = \max \{ 0, 2 / N \} = \frac{2}{N}.
\end{align}

The average concurrence for this state is the same as the minimal one, due to the permutation symmetry that these states present, and it scales as $1/N$:
\begin{align}
\Ccal_{\mathrm{avg}}(\hat\rho_{D})= \Ccal_{\mathrm{min}}(\hat\rho_{D})= \frac{2}{N} .
\end{align}

\begin{figure}
\centering
\includegraphics[scale = 0.2]{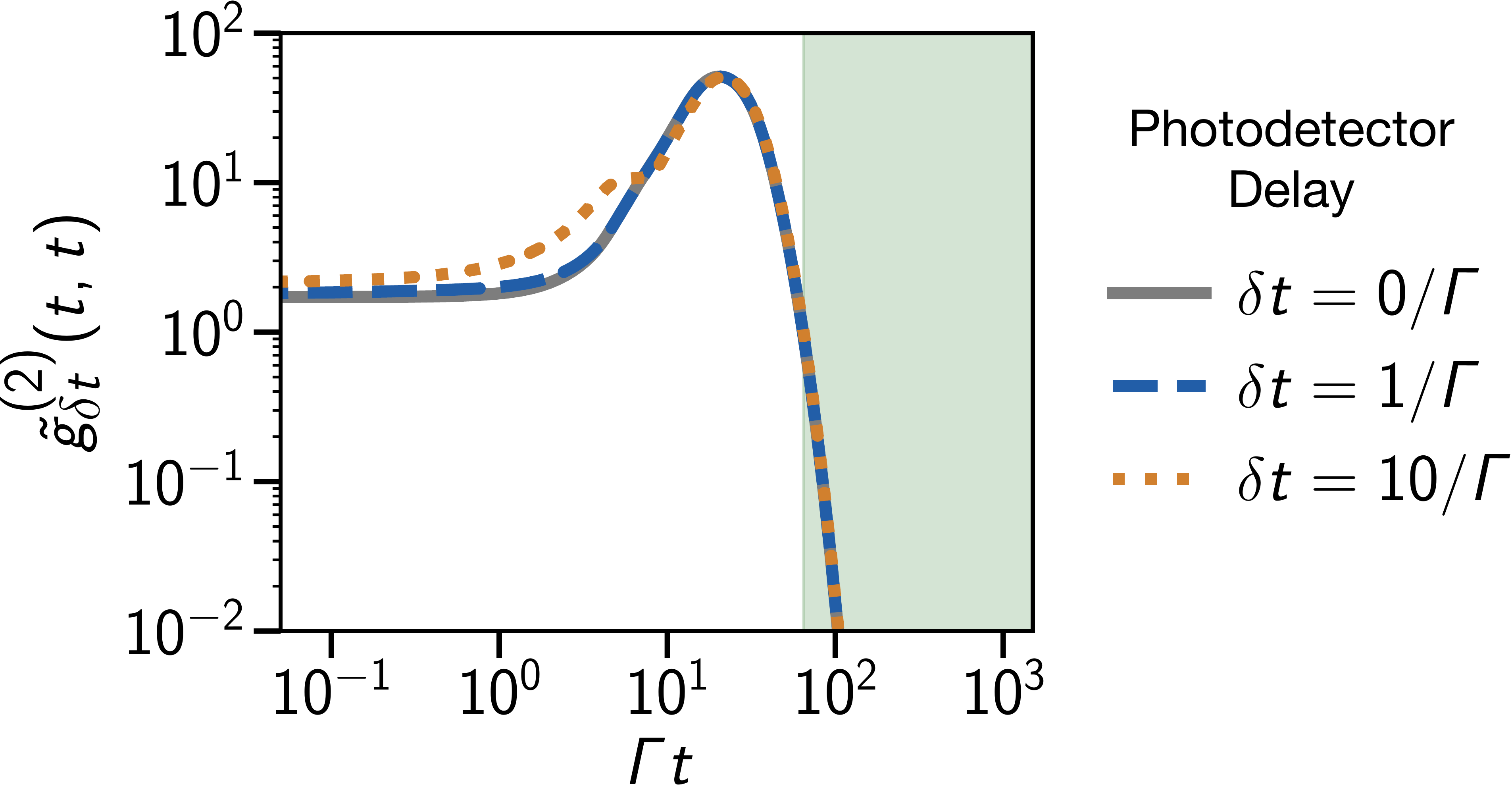}
\caption{Second-order correlation function of the light emitted by a chain of $N=7$ atoms initially in state $\hat{\rho}_{\mathrm{mix}}$, for different time windows for the photodetection. The atomic lattice has a spacing  $d=\pi/2k$.}
\label{Ap-fig:g2delay}
\end{figure}

Let us now consider the case where the system is in a statistical mixture between the single-excitation state $\ket{D_{N}}$ and the ground state $\ket{g_{N}}\!=\!\ket{g}^{\otimes N}$:
\begin{align}
\hat\rho(\epsilon) = (1-\epsilon)\ket{g_{N}}\bra{g_{N}} + \epsilon \ket{D_{N}}\bra{D_{N}} ,
\end{align}
with $0\leq\epsilon\leq 1$. Following the same procedure as before, we compute the reduced density matrix for two atoms and obtain
\begin{align}
\hat\rho_{k\ell}^{D} &= \sum_{j_{r}}^{(k,\ell)} \bra{\{j_{r}\}_{r\neq(k,\ell)}} \left[(1-\epsilon)\ket{g_{N}}\bra{g_{N}}\right] \ket{\{j_{r}\}_{r\neq(k,\ell)}} \nonumber \\
&+ \sum_{j_{r}}^{(k,\ell)} \bra{\{j_{r}\}_{r\neq(k,\ell)}} \left[\epsilon\rho_{D}\right] \ket{\{j_{r}\}_{r\neq(k,\ell)}}. \label{Ap-Eq-RhoklI}
\end{align}
Then, using Eq.~\eqref{Ap-Eq-rholk-Pure} and the fact that $\sum_{j_{r}}^{(k,\ell)} \bra{\{j_{r}\}_{r\neq(k,\ell)}} \left[\ket{g_{N}}\bra{g_{N}}\right] \ket{\{j_{r}\}_{r\neq(k,\ell)}} \!=\!\ket{0_{k\ell}}\bra{0_{k\ell}}$, we get the following expression for the reduced density matrix:
\begin{align}
\hat\rho_{k,\ell}(\epsilon) = \epsilon\hat\rho_{k\ell}^{D} + (1-\epsilon)\ket{0_{k\ell}}\bra{0_{k\ell}}.
\end{align}
The eigenvalues of matrix $\hat R$ for this state are $\lambda_{1}=2\epsilon/N$ and $\lambda_{n\neq1}=0$, so the average and minimum concurrences read
\begin{align}
\Ccal_{\mathrm{avg}}(\hat\rho(\epsilon))= \Ccal_{\mathrm{min}}(\hat\rho_{k,\ell}(\epsilon))= \frac{2\epsilon}{N}.
\end{align}

\section{Finite temporal resolution of the photodetector}

The equal-time correlation function $g^{(2)}(t,t)$ introduced in Eq.~({\color{blue}5}) of the main text corresponds to the cross-measurement $G^{(2)}(t)=\langle \hat{E}^{-}(t)\hat{E}^{-}(t)\hat{E}^{+}(t)\hat{E}^{+}(t)\rangle$, normalized by the intensity $I(t)=\langle \hat{E}^{-}(t)\hat{E}^{+}(t)\rangle$. If we now assume that the photodetectors which compute cross-correlations and the average intensity have a finite temporal resolution $\delta t$, the consequence of this limitation can be evaluated by averaging the cross-correlator and the intensity over this finite time $\delta t$. We thus introduce the average 
\begin{align}
\bar{X}_{\delta t}(t) = \frac{1}{\delta t} \int_{t - \delta t /2}^{t + \delta t /2} d\xi X(\xi),
\end{align}
which results in a second-order correlation function $\tilde{g}^{(2)}_{\delta t}(t,t) = \bar{G}^{(2)}_{\delta t}(t,t)/\bar{I}_{\delta t}^2(t,t)$. 

As can be observed in Fig.~\ref{Ap-fig:g2delay}, while the short-time dynamics presents a $\tilde{g}^{(2)}_{\delta t}(t,t)$ strongly affected by the time window $\delta t$, due to the fast (superradiant) nature of this dynamics, the late dynamics is essentially unaltered by averaging over a window $\delta t=10/\Gamma$.

\begin{figure}
\centering
\includegraphics[scale = 0.43]{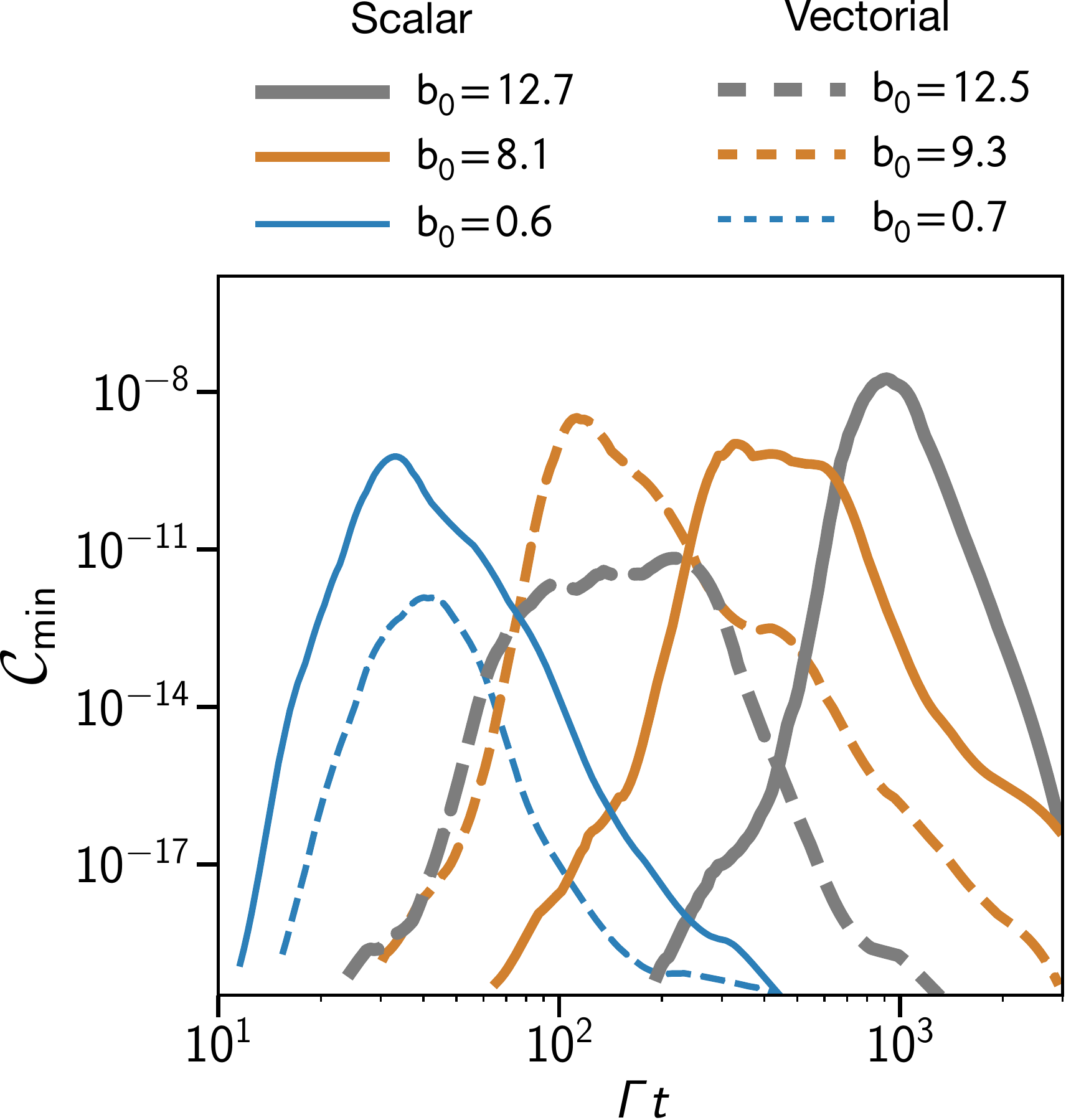}
\caption{Comparison between scalar and vectorial models: Dynamics of the miminum concurrence $\mathcal{C}_\mathrm{min}$, for increasing optical thicknesses $b_0$, and thus increasing atomic densities since the particle number $N=8$ is fixed. The atoms are initially in the statistical mixture~({\color{blue}3}), and the curves were averaged over 10 realizations.}
\label{Ap-fig:scavec}
\end{figure}

\section{Effects of van der Waals Dephasing}

We here compare the increase in lifetime of the global entanglement generated, for the scalar and vectorial model of dipole-dipole interactions. As it can be observed in Fig.~\ref{Ap-fig:scavec}, scalar light is characterized by increasing times for the peak of all-to-all entanglement in the system, as the optical depth (and thus the density $\rho=N/[(4/3)\pi R^3]$, considering that the present simulations are realized with a fixed atom number) is increased. This corresponds to the rise of collective effects as interactions become stronger for scalar-like dipole-dipole interactions, such as proposed in the original work by Dicke~\cite{Dicke1954}.

However, when polarization and near-field terms are accounted for, collective effects are affected by the increasing density, see dashed lines in Fig.~\ref{Ap-fig:scavec}. This effect can be interpreted as a van der Waals dephasing, where the (near-field) non-propagating terms of the dipole-dipole interaction result in a broadening of the system eigenspectrum. This leads to a dephasing over time, which occurs at larger densities for superradiance~\cite{Gross1982}, and at much lower densities for subradiance due to the timescales involved~\cite{Cipris2021b}.

\bibliography{BiblioCollectiveScattering.bib}